\documentclass[12pt]{article}
\usepackage{CJK}
\usepackage{amssymb}
\usepackage{amsmath}
\usepackage{enumerate}
\usepackage{graphicx}
\usepackage{amsfonts}
\usepackage{url}
\usepackage{bm}
\usepackage{tikz}  
\usepackage{pgf}   

\usepackage{pifont}
\usepackage{epsfig,subfigure,dsfont,amsthm,amsbsy,mathrsfs,amscd}
\newtheorem{theorem}{Theorem}

\newtheorem{corollary}{Corollary}

\input amssym.def

\newcommand\btd{\raise 2pt \hbox{$\hat\bigtriangledown$}\hskip 1.5pt}
\newcommand\bt{\raise 2pt \hbox{$\bigtriangledown$}\hskip 1.5pt}
\begin{document}

\title{\large\bf Coherence of Assistance and Assisted Maximally Coherent States}

\author{Ming-Jing Zhao$^1$$^*$, Rajesh Pereira$^2$, Teng Ma$^3$, Shao-Ming Fei$^{4,5}$
\\[10pt]
\footnotesize
\small $^1$School of Science,\\
\small Beijing Information Science and Technology University, Beijing, 100192, P. R. China\\
\footnotesize
\small $^2$ Department of Mathematics and Statistics, \\
\small University of Guelph, Guelph, N1G2W1, Canada\\
\small $^3$ Beijing Academy of Quantum Information Sciences, Beijing 100193, P. R. China\\
\small $^{4}$ School of Mathematical Sciences, \\
\small Capital Normal
University, Beijing, 100048, P. R. China\\
\small $^{5}$ Max-Planck-Institute for Mathematics in the Sciences, 04103
Leipzig, Germany}
\date{}

\maketitle

\centerline{$^*$ Correspondence to zhaomingjingde@126.com}
\bigskip

\begin{abstract}
Coherence and entanglement are fundamental concepts in resource theory. The coherence (entanglement) of assistance is the coherence (entanglement) that can be extracted assisted by another party with local measurement and classical communication. We introduce and study the general coherence of assistance.
First, in terms of real symmetric concave functions on the probability simplex, the coherence of assistance and the entanglement of assistance are shown to be in one-to-one correspondence.
We then introduce two classes of quantum states: the assisted maximally coherent states and the assisted maximally entangled states. They can be transformed into maximally coherent or entangled pure states with the help of another party using local measurement and classical communication.
We give necessary conditions for states to be assisted maximally coherent or assisted maximally entangled. Based on these, a unified framework between coherence and entanglement including coherence (entanglement) measures, coherence (entanglement) of assistance, coherence (entanglement) resources is proposed.
Then we show that the coherence of assistance as well as entanglement of assistance are strictly larger than the coherence of convex roof and entanglement of convex roof for all full rank density matrices. So all full rank quantum states are distillable in the assisted coherence distillation.
\end{abstract}


Quantum coherence is an important feature in quantum physics and is of practical significance in quantum computation and quantum communication \cite{A. Streltsov-rev,M. Hu}.
The formulation of the resource theory of coherence was initiated in Ref. \cite{T. Baumgratz}, in which some requirements are proposed for a well defined quantum coherence measure. Later on, coherence measures or monotones such as
the $l_1$ norm of coherence \cite{T. Baumgratz}, the relative entropy of coherence \cite{T. Baumgratz}, intrinsic randomness of coherence \cite{X. Yuan},  coherence concurrence \cite{X. Qi}, distillable coherence \cite{A. Winter}, coherence cost \cite{A. Winter},
robustness of coherence \cite{C. Napoli}, coherence number \cite{S. Chin} and so on are proposed theoretically or operationally.
Many of these coherence measures were either created from entanglement measures using a standard modification or are closely related to one that was.
For example, the
robustness of coherence and coherence number are defined in a manner similar to that of robustness of entanglement and the Schmidt
number in entanglement theory respectively \cite{G. Vidal1999,B. M. Terhal}. The $l_1$ norm coherence is exactly the twice negativity for pure states \cite{S. Rana}.

Coherence of assistance is another quantifier which quantifies the coherence that can be extracted assisted by another party under local measurements and classical communication \cite{E. Chitambar}.
Suppose Alice holds a state $\rho^A=\sum_k p_k |\psi_k\rangle\langle \psi_k|$ with coherence $C(\rho^A)$. Bob holds another part of the purified state of $\rho^A$. The joint state between Alice and Bob is $\sum_k p_k |\psi_k\rangle_A \otimes |k\rangle_B$. Bob performs local measurements $\{|k\rangle\langle k|\}$ and informs Alice the measurement outcomes by classical communication. Alice's quantum state will be in a pure state ensemble $\{ p_k,\  |\psi_k\rangle\langle \psi_k|\}$ with average coherence $\sum_k p_k C(|\psi_k\rangle\langle \psi_k|)$. The process is called assisted coherence distillation. The maximum average coherence is called the coherence of assistance which quantifies the one-way coherence
distillation rate \cite{E. Chitambar}. The coherence of assistance is always greater than or equal to the coherence measure. But it is still not clear whether one can always obtain more coherence with the help of another party. Our answer in this paper is that one can always obtain more coherence for the full ranked quantum states.

As with other measures of coherence and entanglement, the coherence of assistance and the entanglement of assistance are also closely related. In fact, the relative entropy coherence of assistance is equal to the entanglement formation of assistance \cite{E. Chitambar,D. DiVincenzo,O. Cohen,J. A. Smolin} and the $l_1$ norm coherence of assistance corresponds to the convex-roof extended negativity entanglement of assistance \cite{zhao l1max,J. S. Kim,G. Vidal}.
As intrinsic characteristics of quantum physics, the inextricable relationship between quantum coherence and quantum entanglement is not limited to specific quantum coherence measures and entanglement measures as well as the coherence of assistance and the entanglement of assistance. Ref. \cite{A. Streltsov} shows any coherence can be converted to entanglement via incoherent operations, and each entanglement measure corresponds to a coherence measure. It has been further shown that coherence can be converted to bipartite nonlocality, genuine tripartite entanglement and genuine tripartite nonlocality \cite{xiya}.
In Refs. \cite{K. C. Tan2016,K. C. Tan2018} the authors construct an entanglement monotone based on any given coherence measure. More generally, Ref.\ \cite{H. Zhu} establishes a general operational one-to-one mapping between coherence measures and entanglement measures.

Inspired by these results, we aim to construct a general relation between the coherence of assistance and entanglement of assistance for the coherence and entanglement theory. First we review the construction of entanglement measures and coherence measures using the convex roof extension. Then we define the general coherence of assistance and the one-to-one correspondence between entanglement of assistance and coherence of assistance is established afterwards.
Subsequently, two special classes of states called assisted maximally coherent states and assisted maximally entangled states are introduced. These states can be turned into the maximally coherent or maximally entangled states with the help of another party's local measurement and classical communication. The necessary conditions for states to be the assisted maximally coherent states or assisted maximally entangled states are presented. These states can be thought of as potentially perfect coherence or entanglement resources.
Then we show the coherence of convex roof and the coherence of assistance, as well as the entanglement measure in convex roof construction
and the entanglement of assistance, are not equal for any full rank density matrix.
This demonstrates that these kind of states are all distillable in the assisted coherence distillation.
The unified framework between coherence and entanglement is shown in Fig.1.

\begin{figure}

\begin{tikzpicture}[->=stealth,auto,node distance=2cm]
\centering
\node (f) 
{$f$};

\node (ef) [xshift=0.3cm,right of =f] 
{$E_f$};

\node (cf) [xshift=-0.3cm,left of =f] 
{$C_f$};

\node (ec) [xshift=0.3cm,yshift=1cm,right of =ef] 
{$E_c$};

\node (cc) [xshift=-0.3cm,yshift=1cm,left of =cf] 
{$C_c$};

\node (ca) [xshift=-0.3cm,yshift=-1cm,left of=cf] 
{$C_a$};

\node (ea) [xshift=0.3cm,yshift=-1cm,right of=ef] 
{$E_a$};
\node (me) [xshift=0.3cm,right of=ec] {ME}; 
\node (mc) [xshift=-0.3cm,left of=cc]  {MC};
\node (ame) [xshift=0.3cm,right of=ea] {AME};
\node (amc) [xshift=-0.3cm,left of=ca]  {AMC};

\draw[<->] (f) -- (cf) ;  
\draw[<->] (f) -- (ef) ; 
\draw[<->] (cf) -- (cc) node[midway,above,sloped]{};
\draw[<->] (ef) -- (ec) node[midway,above,sloped]{};
\draw[<->] (cf) -- (ca) node[midway,below,sloped]{};
\draw[<->] (ef) -- (ea) node[midway,below,sloped]{};
\draw[->,dashed] (ec) -- (me);
\draw[->,dashed] (ea) -- (ame) ; 
\draw[->,dashed] (cc) -- (mc);
\draw[->,dashed] (ca) -- (amc) ; 
\draw (1.2,-2.5) rectangle (8,2) node[midway,below=3cm]{Entanglement};
\draw (-8,-2.5) rectangle (-1.2,2) node[midway,below=3cm]{Coherence};
\end{tikzpicture}
\caption{Relations between coherence and entanglement.
Here $f\in \mathfrak{F}\setminus \{0\}$.
$E_f$ is a function defined on bipartite pure states as in Eq. (\ref{ef}).
$C_f$ is a function defined on pure states as in Eq. (\ref{cf}).
$E_c$ is the entanglement measure called the entanglement of convex roof which is the convex roof extension of $E_f$ from pure states to mixed states in Eq. (\ref{min ef}).
$C_c$ is the coherence measure called the coherence of convex roof which is the convex roof extension of $C_f$ from pure states to mixed states in Eq. (\ref{min cf}).
$E_c$ and $C_c$ are one-to-one corresponded by the real symmetric concave function $f$.
The maximum points of $E_c$ and $C_c$ are maximally entangled states (ME) in Eq. (\ref{max ent pure}) and maximally coherent states (MC) in Eq. (\ref{MC}) respectively.
$E_a$ is the entanglement of assistance which is the least concave majorant extension of $E_f$ from pure states to mixed states in Eq. (\ref{Ea}).
$C_a$ is the coherence of assistance which is the least concave majorant extension of $C_f$ from pure states to mixed states in Eq. (\ref{Ca}).
$E_a$ and $C_a$ are one-to-one corresponded by the real symmetric concave function $f$.
The maximum points of $E_a$ and $C_a$ are assisted maximally entangled states (AME) in definition 2 and assisted maximally coherent states (AMC) in definition 1 respectively. The coherence of assistance $C_a$ as well as the entanglement of assistance $E_a$ is shown to be strictly larger than the coherence of convex roof $C_c$ and entanglement of convex roof $E_c$ for all full rank quantum states.
}
\end{figure}
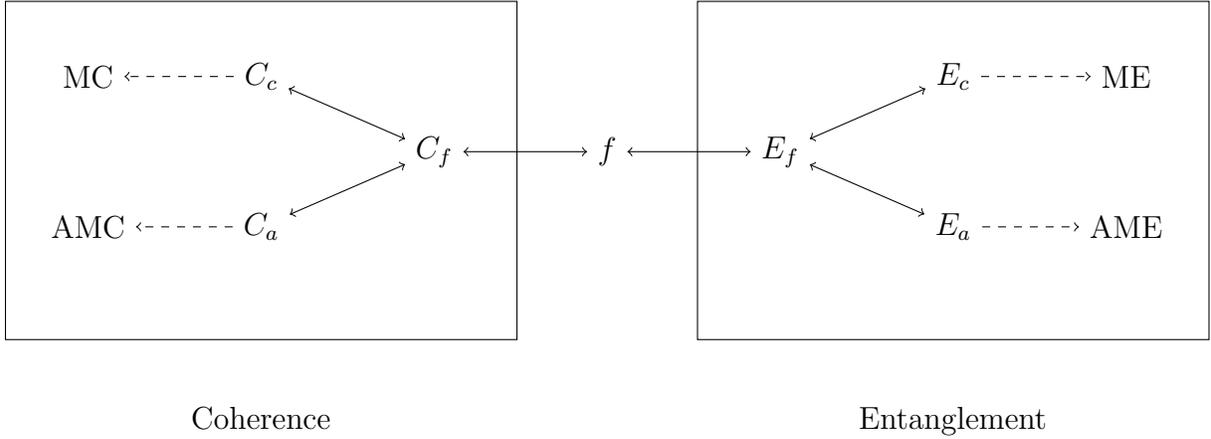

\medskip
\noindent{\bf Results}

{\bf Entanglement of assistance and coherence of assistance}
A state is called incoherent if the density matrix $\rho$ is diagonal in the fixed reference basis $\{|i\rangle\}$, $\rho=\sum_i p_i |i\rangle\langle i|$ with $p_i$ the probabilities. Otherwise the state is called coherent. Obviously, coherence is basis dependent. A completely positive trace preserving map $\Lambda$ acting as $\Lambda(\rho)=\sum_l K_l\rho K_l^\dagger$ is said to be an incoherent operation if all the Kraus operators $K_l$ map incoherent states to incoherent states \cite{T. Baumgratz}. A coherence measure $C(\rho)$ should satisfy  \cite{T. Baumgratz}: (1) $C(\rho)\geq 0$ with $C(\rho)=0$ if and only if $\rho$ is incoherent. (2) $C(\rho)$ is nonincreasing under incoherent operations $\Lambda$, $C(\rho)\geq C(\Lambda(\rho))$. (3) $C(\rho)$ is nonincreasing on average under selective incoherent operations, $C(\rho)\geq \sum_l q_l C(\rho_l)$, with $q_l=tr(K_l \rho K_l^\dagger)$ and $\rho_l=K_l \rho K_l^\dagger/q_l$. (4) $C(\rho)$ is a convex function on the density matrices, $C(\sum_j p_j \rho_j)\leq \sum_j p_j C(  \rho_j)$.

Let ${\mathfrak{{F}}}=\{f\}$ be the set of functions on the probability simplex $\Omega=\{\vec{x}=(x_0,x_1,\cdots,x_{n-1})^T|\sum_{i=0}^{n-1} x_i =1 \ \text {and} \ x_i\geq0 \}$ such that (i) $f$ is a real symmetric concave function; (ii) $f((1,0,\cdots,0)^T)=0$.
We assume $f\in {\mathfrak{{F}}}\setminus \{0\}$ in this paper. Under these conditions $f$ yields an entanglement monotone $E_f$ for the $n\otimes n$ pure states. If $|\psi\rangle$ has the Schmidt form $|\psi\rangle=\sum_{i=0}^{n-1} \lambda_i |i_A \rangle |i_B\rangle$ with $\lambda_i\geq 0$, $E_f$ can be defined as
\begin{equation}\label{ef}
E_f(|\psi\rangle)=f(\lambda(|\psi\rangle)),
\end{equation}
where $\{|i_A \rangle\}_{i=0}^{n-1}$ and $\{|i_B \rangle\}_{i=0}^{n-1}$ are orthonormal basis of the two subsystems, and $\lambda(|\psi\rangle)=(\lambda_0^2, \lambda_1^2, \cdots, \lambda_{n-1}^2)^T$.
The entanglement monotone $E_f$ can be extended to mixed states by the convex roof construction \cite{H. Zhu,G. Vidal2000}. The entanglement of convex roof $E_{c}$ is given by
\begin{equation}\label{min ef}
E_{c}(\rho)=\min \sum_k p_k {E_f}(|\psi_k\rangle),
\end{equation}
where $E_f$ is defined by (\ref{ef}), the minimization is taken over all pure state decompositions of $\rho=\sum_k p_k |\psi_k\rangle\langle\psi_k|$.

The entanglement of convex roof $E_{c}$ is an entanglement measure \cite{G. Vidal2000}. Any quantum states that are local unitary equivalent to
\begin{equation}\label{max ent pure}
|\phi^+\rangle=\frac{1}{\sqrt{n}} \sum_{j=0}^{n-1} |jj\rangle,
\end{equation}
are all maximally entangled according to $E_{c}$. These states are the only ones such that $E_{c}$ attains its maximum for $n \otimes n$ systems.

Correspondingly, for the fixed reference basis $\{|i\rangle\}$ and any $f\in {\mathfrak{{F}}}\setminus \{0\}$, a coherence measure for pure state $|\psi\rangle=\sum_{i=0}^{n-1} \psi_i |i\rangle$ can be defined as
\begin{equation}\label{cf}
C_f(|\psi\rangle)=f(\mu(|\psi\rangle)),
\end{equation}
where $\mu(|\psi\rangle)=(|\psi_0|^2, |\psi_1|^2, \cdots, |\psi_{n-1}|^2)^T$ is the coherence vector. The coherence measure $C_f$ can be extended to mixed states by the convex roof construction \cite{H. Zhu,S. Du}. The coherence of convex roof $C_{c}$ is given by
\begin{equation}\label{min cf}
C_{c}(\rho)=\min \sum_k p_k {C_f}(|\psi_k\rangle),
\end{equation}
where $C_f$ is defined by (\ref{cf}), the minimization is taken over all pure state decompositions of $\rho=\sum_k p_k |\psi_k\rangle\langle\psi_k|$.

The coherence of convex roof $C_{c}$ is a coherence measure \cite{A. Streltsov-rev, M. Hu, T. Baumgratz, H. Zhu}. According to
the coherence measure $C_{c}$, all maximally coherent states in an $n$-dimensional system can be transformed into the pure states in the following set by unitary incoherent operations \cite{Y. Peng}:
\begin{equation}\label{MC}
\{\frac{1}{\sqrt{n}} \sum_{j=0}^{n-1} e^{{\rm i}\theta_j} |j\rangle\ |\  \theta_1, \cdots,  \theta_{n-1}\in [0,\ 2\pi)\}.
\end{equation}

For any function $f\in \mathfrak{F}\setminus \{0\}$, the entanglement monotone $E_f$ can be also extended to mixed states by the least concave majorant extension, giving rise to entanglement of assistance.
The entanglement of assistance can be defined by
\begin{equation}\label{Ea}
E_a(\rho)=\max \sum_k p_k E_f(|\psi_k\rangle),
\end{equation}
where the maximization is taken over all pure state decompositions of $\rho=\sum_k p_k |\psi_k\rangle\langle\psi_k|$.

The entanglement of assistance has been introduced with respect to some specific functions \cite{D. DiVincenzo,O. Cohen,J. A. Smolin,J. S. Kim}. Definition (\ref{Ea}) presents a general notion of entanglement of assistance for arbitrary function $f\in \mathfrak{F}\setminus \{0\}$.
It is a dual construction to the entanglement of convex roof. Unlike the entanglement of convex roof which is an entanglement measure, the entanglement of assistance is not a measure of entanglement, as it is not monotonic under local operations and classical communications \cite{G. Gour}. But the entanglement of assistance describes the hidden entanglement that can be unlocked with the help of another party's local measurement and classical communication.

Correspondingly, we can define the coherence of assistance,
\begin{equation}\label{Ca}
C_a(\rho)=\max \sum_k p_k C_f(|\psi_k\rangle),
\end{equation}
with $C_f$ defined in Eq. (\ref{cf}), where the maximization is taken over all pure state decompositions of $\rho=\sum_k p_k |\psi_k\rangle\langle\psi_k|$.

We observe that $C_a$ vanishes if the quantum state is incoherent and pure.
Additionally, $C_a$ is not monotonic under incoherent operations. For example, consider $\rho=|0\rangle\langle0|$ and an incoherent operation $\Lambda(\rho)=K_1\rho K_1^\dagger +  K_2\rho K_2^\dagger$, where $K_1=\frac{1}{\sqrt{2}}I$ and $K_2=\frac{1}{\sqrt{2}}(|0\rangle\langle1|+|1\rangle\langle0|)$ satisfying $K_1^\dagger K_1 +  K_2^\dagger K_2=I$. After the incoherent operation, $\Lambda(\rho)=\frac{1}{2}(|0\rangle\langle0|+|1\rangle\langle1|)=\frac{1}{2}(|\psi_1\rangle\langle\psi_1|+|\psi_2\rangle\langle\psi_2|)$, with $|\psi_1\rangle=\cos \theta|0\rangle+ \sin \theta|1\rangle$ and $|\psi_2\rangle=-\sin \theta|0\rangle + \cos \theta|1\rangle$. By the assumptions of $f$ we know that there exists an angle $\theta$ such that $C_f(|\psi_1\rangle)=C_f(|\psi_2\rangle)>0$. Hence,
$0=C_a(\rho)< C_a(\Lambda(\rho))$, which violates the monotonicity of coherence measures under incoherent operations. Therefore, the coherence of assistance is actually not a coherence measure.

\begin{theorem}\label{th correspondence}
The coherence of assistance ${C_a}$ corresponds one-to-one to the entanglement of assistance $E_a$.
\end{theorem}

See the methods section for the proof of the Theorem \ref{th correspondence}.

Under the product reference bases, the entanglement of assistance $E_a$ is just the coherence of assistance $C_a$ for all pure states as well as for Schmidt correlated states $\rho_{mc}=\sum_{ij} \rho_{ij}|ii\rangle\langle jj|$ \cite{E. Rains}, $E_a(\rho_{mc})=C_a(\rho_{mc})$. The similar results also hold true for the entanglement of convex roof $E_c$ and the coherence of convex roof $C_c$. The correspondence not only bridges coherence theory and entanglement theory, but also generalizes many results in entanglement theory to coherence theory.

The entanglement of assistance $E_a$ and coherence of assistance $C_a$ depend on the choice of the functions $f\in\mathfrak{F}\setminus \{0\}$.
If $f(\vec{p})=-\sum_i p_i \log p_i$ for $\vec{p}=(p_1,p_2,\cdots,p_n)^T$ in the probability simplex, $E_a$ becomes the entanglement of formation of assistance \cite{D. DiVincenzo,O. Cohen,J. A. Smolin}, and $C_a$ becomes the relative entropy coherence of assistance \cite{E. Chitambar}.
If $f(\vec{p})=\sum_{i\neq j}\sqrt{p_ip_j}$ for $\vec{p}=(p_1,p_2,\cdots,p_n)^T$ in the probability simplex, then $E_a$ becomes the half convex-roof
extended negativity of assistance \cite{J. S. Kim} and $C_a$ becomes the $l_1$ norm coherence of assistance \cite{zhao l1max}. Analogously, one can also define various other types of entanglement of assistance and coherence of assistance based on other real symmetric concave functions $f$.
For example, let $f(\vec{p})=\sqrt{2(1-\sum_i p_i^2)}$, then $E_a$ is the entanglement of assistance in terms of concurrence \cite{Z. G. Li}, in which an upper bound of entanglement of assistance is provided as $E_a(\rho)\leq \sqrt{2(1-tr(\rho_A^2))}$ with $\rho_A=tr_B (\rho)$. For this function $f$, we can define the coherence of assistance $C_a$ in terms of concurrence similarly and one upper bound is $C_a(\rho)\leq \sqrt{2(1-\sum_i \rho_{ii}^2)}$ with $\rho_{ii}$ the diagonal entries of $\rho$ in the reference basis.

{\bf Assisted maximally coherent states and assisted maximally entangled states}
The average of entanglement and coherence depends on the ensembles of a quantum state.
Assisted by another party, the entanglement of assistance and coherent of assistance attain the maximum average entanglement and coherence of the quantum state. Here we investigate two classes of states called assisted maximally coherent states and assisted maximally entangled states for which the maximal average coherence and entanglement are the same as the maximally coherent states and maximally entangled states.

{\sf [Definition 1]} We call an $n$ dimensional quantum state $\rho$ assisted maximally coherent (AMC) if it is a convex combination of maximally coherent pure states.

The AMC states are a class of states that achieve the maximum of coherence of assistance. Therefore they are a potentially perfect coherence resource.
For pure states, all the maximally coherent states are AMC and vice versa. For mixed states, all maximally mixed states $\rho=\frac{1}{n} \sum_{i=0}^{n-1} |i\rangle\langle i|$ are AMC. This follows from the existence of a maximally coherent pure state decomposition $\{p_k,\  |\psi_k\rangle\}$ of $\rho$, where $p_k=\frac{1}{n}$ for all $k$ and $|\psi_k\rangle= \frac{1}{\sqrt{n}} \sum_{j=0}^{n-1} e^{2\pi {\rm i} (k-1)j/n}|j\rangle$ for $k=1,2,\cdots,n$, and $\rm i=\sqrt{-1}$ is the imaginary unit. The Fourier matrix $F$ with its $k$-th column given by the vector ${\sqrt{n}}|\psi_k\rangle$ satisfies $FF^\dagger=nI$ and $|F_{kj}|=1$, $k=1,\cdots, n$; $j=0,1,2,\cdots,n-1$. Therefore, $\{|\psi_k\rangle\}_{k=1}^n$ is an orthonormal basis of the $n$ dimensional system, which means that $ \sum_{i=0}^{n-1} |i\rangle\langle i|= \sum_{k=1}^n |\psi_k\rangle\langle\psi_k|$.

\begin{theorem}\label{cor nec con for AMC}
If an $n$ dimensional quantum state $\rho=\sum_{ij} \rho_{ij}|i\rangle \langle j|$ is AMC, then $\rho_{ii}=\frac{1}{n}$ for all $i$, which becomes both necessary and sufficient for two and three dimensional systems.
\end{theorem}

See the methods section for the proof of the Theorem \ref{cor nec con for AMC}.

There exist $n$-dimensional quantum states $\rho$ with all diagonal entries $\frac{1}{n}$ which do not allow for pure state decomposition $\{p_k,\  |\psi_k\rangle\}$ such that all diagonal entries of $|\psi_k\rangle\langle \psi_k|$ are $\frac{1}{n}$ for all $k$ and $n\geq 4$. Some specific examples are shown in Refs. \cite{B. Regula,U. Haagerup}. We now give an explicit pure state decomposition for three dimensional AMC states. In a three dimensional system, the quantum state $\rho=\sum_{i,j} \rho_{ij} |i\rangle\langle j|$, with $\rho_{11}=\rho_{22}=\rho_{33}=\frac{1}{3}$ and real nonzero off diagonal entries, is an example of mixed AMC state that is not a maximally mixed state. Let $p_1=\frac{1}{4}(1+\rho_{12} +\rho_{13} +\rho_{23} )$, $p_2=\frac{1}{4}(1-\rho_{12} -\rho_{13} +\rho_{23} )$, $p_3=\frac{1}{4}(1-\rho_{12} +\rho_{13} -\rho_{23} )$, $p_4=\frac{1}{4}(1+\rho_{12} -\rho_{13} -\rho_{23} )$, and
$|\psi_1\rangle=\frac{1}{\sqrt{3}}(|1\rangle+|2\rangle+|3\rangle)$, $|\psi_2\rangle=\frac{1}{\sqrt{3}}(-|1\rangle+|2\rangle+|3\rangle)$,
$|\psi_3\rangle=\frac{1}{\sqrt{3}}(|1\rangle-|2\rangle+|3\rangle)$,
$|\psi_4\rangle=\frac{1}{\sqrt{3}}(|1\rangle+|2\rangle-|3\rangle)$,
then $\{p_k,\ |\psi_k\rangle\}$ is a pure state decomposition of $\rho$ with components all maximally coherent.

Similar to AMC states, we can define the assisted maximally entangled (AME) states in bipartite systems.

{\sf [Definition 2]} An $n\otimes n$ bipartite quantum state $\rho$ is called assisted maximally entangled (AME) if it is a convex combination of maximally entangled pure states.

\begin{theorem}\label{th AME}
The $n\otimes n$ Schmidt correlated state $\rho_{mc}=\sum_{ij} \rho_{ij}|ii\rangle\langle jj|$ is AME if and only if the $n$ dimensional state $\rho=\sum_{ij} \rho_{ij}|i\rangle\langle j|$ is AMC.
\end{theorem}

See the methods section for the proof of the Theorem \ref{th AME}.
Combining Theorems \ref{cor nec con for AMC} and \ref{th AME}, we get the following necessary condition for Schmidt correlated states to be AME.

\begin{corollary}\label{cor ame}
If an $n\otimes n$ Schmidt correlated state $\rho_{mc}=\sum_{ij} \rho_{ij}|ii\rangle \langle jj|$ is AME, then $\rho_{ii}=\frac{1}{n}$ for all $i$, which is both sufficient and necessary for the cases of $n=2$ and $n=3$ systems.
\end{corollary}

For pure states, all maximally entangled states are AME and vice versa. For mixed states, all maximally correlated states $\rho=\frac{1}{n} \sum_{i=0}^{n-1} |ii\rangle\langle ii|$ are AME due to Corollary \ref{cor ame}. Besides the Schmidt correlated states, there are also other AME states.
As examples, consider two-qubit system. Let $\rho=p|\psi_1\rangle\langle\psi_1|+(1-p)|\psi_2\rangle\langle\psi_2|$ with $0<p<1$, $|\psi_1\rangle=\frac{1}{\sqrt{2}}(|00\rangle+|11\rangle)$ and $|\psi_2\rangle=\frac{1}{\sqrt{2}}(|01\rangle+|10\rangle)$. Clearly, $\rho$ is AME but not Schmidt correlated. The maximally mixed states $\rho=\frac{1}{n^2} \sum_{i,j=0}^{n-1} |ij\rangle\langle ij|$ are AME, since they can be written as the average of generalized Bell states $|\phi_{st}\rangle=I\otimes U_{st}^* |\phi^+\rangle$, where $U_{st}=h^tg^s$, $h|j\rangle=|j+1\mod n\rangle$, $g|j\rangle=\omega^j |j\rangle$, $\omega=\exp(-2\pi {\rm i}/n)$, and superscript $*$ stands for the conjugate \cite{S. Albeverio}.

AMC states and AME states are potential maximally coherent states and maximally entangled states, since they can be decomposed as the convex combinations of maximally coherent and maximally entangled pure states, respectively. Furthermore, they can be collapsed to maximally coherent states and maximally entangled states with the help of another party's local measurements and classical communication operationally, if only one knows the optimal pure state decompositions. As applications, one can transform the AMC states to maximally coherent pure states with the help of another party's local measurements and classical communication for the purpose of quantum information processing such as the Deutsch-Jozsa algorithm to speedup the computation \cite{M. Hillery}. In this sense, the AMC states are potentially perfect quantum resources. In fact, the experimental realization in linear optical systems for obtaining the coherence of assistance with respect to the relative entropy coherence in two dimensional systems has already been presented \cite{K. D. Wu}.

{\bf Relation between the convex roof extension and the least concave majorant extension}
The strict relation between the coherence of convex roof and the coherence of assistance, that is, whether $C_c(\rho)< C_a(\rho)$ holds for all mixed quantum states is an interesting topic. The physical motivation is from the coherence distillation, which is to extract pure coherence from a mixed state by incoherent operations \cite{A. Winter}. All coherent states can be distilled by the coherence distillation process.
The assisted coherence distillation is then introduced to generate the maximal possible coherence with the help of another party's local measurements and classical communication \cite{E. Chitambar}. The relative entropy coherence of assistance in form of Eq. (\ref{Ca}) with a specific function $f$ is proposed first there to quantify the one way coherence
distillation rate in the assisted coherence distillation. Generally we can get more coherence in the assisted coherence distillation. But a natural question is whether we can extract more coherence from all mixed states in the assisted coherence distillation. This question is factually equivalent to whether the coherence of assistance is strictly larger than the coherence of convex roof for all mixed quantum states. If it is true, all mixed quantum states are distillable in the assisted coherence distillation process. In order to answer this question, we consider a much more general case as follows.

We now investigate  the general relations between the convex roof extension and the least concave majorant extension of an arbitrary nonnegative function. Let $\mathcal{H}$ be a finite-dimensional Hilbert space and $F$ a nonnegative function defined on the pure states of $\mathcal{H}$. Define $F_a(\rho)=\max \sum_k p_k F(|\psi_k\rangle)$ to be the least concave majorant extension from $F$,
and $F_c(\rho)=\min \sum_k p_k F(|\psi_k\rangle)$ the convex roof extension from $F$,
where the maximization and minimization are both taken over all pure state decompositions of $\rho=\sum_k p_k |\psi_k\rangle\langle\psi_k|$, respectively.
The convex roof extension $F_c$ is the largest convex function which is equal to $F$ on the pure states while the least concave majorant extension $F_a$ is the smallest concave function. The definitions $F_c$ and $F_a$ are more general than $E_c$, $C_c$ and $E_a$, $C_a$.

\begin{theorem}\label{th ca=cc}
Let $\mathcal{H}$ be a finite-dimensional Hilbert space and $F$ a nonnegative function defined on the pure states in $\mathcal{H}$.
Let $\rho_o$ be a density matrix on $\mathcal{H}$ and $R(\rho_o)$ the range of $\rho_o$. If $F_a(\rho_o)=F_c(\rho_o)$, then there exists a positive semidefinite operator $Q$ on $R(\rho_o)$ such that $F(|\psi\rangle)=\langle \psi |Q| \psi \rangle$ for all pure states $|\psi\rangle$ in $R(\rho_o)$.
\end{theorem}

See the methods section for the proof of the Theorem \ref{th ca=cc}.
Theorem \ref{th ca=cc} transforms the equation $F_a(\rho)=F_c(\rho)$  into the existence of a positive semidefinite operator $Q$ on $R(\rho)$.
So in order to check the coincidence of $F_a(\rho)=F_c(\rho)$, one only needs to check the existence of $Q$ for all pure states in the support of $R(\rho)$.
We apply Theorem \ref{th ca=cc} to coherence theory and entanglement theory.

\begin{corollary}\label{cor coh}
For full rank quantum states $\rho$, the coherence of assistance is strictly larger than the coherence of convex roof, $C_c(\rho)< C_a(\rho)$.
\end{corollary}

\begin{corollary}\label{cor ent}
For full rank bipartite quantum states $\rho$, the entanglement of assistance is strictly larger than the entanglement of convex roof, $E_c(\rho)< E_a(\rho)$.
\end{corollary}

See the methods section for the proof of the corollary \ref{cor coh}.
The proof of Corollary \ref{cor ent} is similar to that of Corollary \ref{cor coh}.
Combined with the physical explanation of coherence (entanglement) of assistance, Corollary \ref{cor coh} and \ref{cor ent} demonstrate that for full rank density matrices, their coherence (entanglement) can be strictly increased with the help of another party's local measurements and classical communication. Hence, this kind of states are distillable in the assisted coherence (entanglement) distillation.

\medskip
\noindent{\bf Discussions}

We have introduced the general coherence of assistance in terms of real symmetric concave functions on the probability simplex, the coherence of assistance and the entanglement of assistance are shown to be in one-to-one correspondence as entanglement measures and coherence measures in the convex roof construction.
Assisted maximally coherent states and assisted maximally entangled states are proposed as the convex combination of the maximally coherent states and maximally entangled states respectively, which can act potentially as perfect resource in quantum information. A necessary and sufficient condition for two or three-dimensional states to be AMC or AME is presented.
Moreover, we have shown that
the coherence of convex roof and the coherence of assistance are not equal for any full rank density matrix, together with a similar result for the entanglement with convex roof construction and the entanglement of assistance.
These results may help strengthen our understanding of the important resources quantum coherence and entanglement.

\medskip
\noindent{\bf Methods}

{\sf Proof of Theorem \ref{th correspondence}}
The coherence measure $C_f$ in Eq. (\ref{cf}) corresponds one-to-one to the real symmetric concave function $f\in {\mathfrak{{F}}}\setminus \{0\}$ \cite{H. Zhu,S. Du}. The entanglement measure $E_f$ in Eq. (\ref{ef}) also corresponds one-to-one to the real symmetric concave function $f\in {\mathfrak{{F}}}\setminus \{0\}$ \cite{G. Vidal2000}. Therefore, the coherence measure $C_f$ in Eq. (\ref{cf}) and the entanglement measure $E_f$ in Eq. (\ref{ef}) are in one-to-one correspondence. As the least concave majorant extension of the coherence measure $C_f$ and entanglement measure $E_f$, the coherence of assistance ${C_a}$ and the entanglement of assistance $E_a$ are also in one-to-one correspondence.

{\sf Proof of Theorem \ref{cor nec con for AMC}}
Before we prove the theorem, we first introduce the concept of a correlation matrix \cite{C. K. Li}. An $n\times n$ Hermitian matrix is called a correlation matrix if it is a positive semidefinite matrix with all diagonal entries being 1. The set of correlation matrices is compact and convex. The extreme points of the set are the correlation matrices with rank 1 for $n=2,3$ \cite{Christensen}. (For $n\ge 4$, there are extreme $n$ by $n$ correlation matrices which have rank two). Hence all $n\times n$ correlation matrices can always be decomposed into the convex combination of rank 1 correlation matrices for $n=2,3$.

Since the diagonal entries of density matrices of maximally coherent pure states are all equal to $\frac{1}{n}$, as the convex combination of maximally coherent pure states, the diagonal entries of AMC states are $\rho_{ii}=\frac{1}{n}$ for all $i$. Therefore, all $n$ dimensional AMC states are the correlation matrices scaled by a multiplicative factor of $\frac{1}{n}$. Since the maximally coherent pure states correspond to the rank 1 correlation matrices, and all $2\times 2$ and $3\times 3$ correlation matrices can be decomposed into a convex combination of rank 1 correlation matrices, the AMC states correspond exactly to the set of correlation matrices for $n=2,3$.  (For $n\geq 4$, the AMC states correspond to a proper subset of the $n\times n$ correlation matrices).
This implies that all diagonal entries being equal to $\frac{1}{n}$ is necessary and sufficient for two and three dimensional AMC states.

{\sf Proof of Theorem \ref{th AME}}
Note that the pure state decompositions of the Schmidt correlated state $\rho_{mc}$ are all of the Schmidt form $|\psi^\prime\rangle=\sum_i a_i|ii\rangle$ \cite{M. J. Zhao2008}. Then $\{ p_k,\  |\psi^{\prime}_k\rangle\}$ is a pure state decomposition of $\rho_{mc}$ with $|\psi_k^\prime\rangle=\sum_i a_i^{(k)}|ii\rangle$ if and only if $\{ p_k,\  |\psi_k\rangle\}$ is a pure state decomposition of $\rho$ with $|\psi_k\rangle=\sum_i a_i^{(k)}|i\rangle$. While $\sum_i a_i^{(k)}|ii\rangle$ is maximally entangled if and only if $\sum_i a_i^{(k)}|i\rangle$ is maximally coherent. Therefore, $\rho_{mc}=\sum_{ij} \rho_{ij}|ii\rangle\langle jj|$ is AME if and only if $\rho=\sum_{ij} \rho_{ij}|i\rangle\langle j|$ is AMC.

{\sf Proof of Theorem \ref{th ca=cc}}
For all $0\leq \tau\leq \rho_o$, define $\tilde{F}(\tau)$ as
\begin{equation}\label{r1}
\tilde{F}(\tau)=\sum_k q_k F(|\psi_k\rangle),
\end{equation}
where $\sum_k q_k |\psi_k\rangle\langle\psi_k|$ is any pure state decomposition of $\tau$ into a weighted sum of pure states,
i.e. $q_k\geq 0$ for all $k$ and $\sum_k q_k\leq 1$. We claim that $\tilde{F}(\tau)$ does not depend on the pure state decomposition at hand.
Indeed, let $\sum_{k^\prime} q_{k^\prime} |\psi^\prime_{k^\prime}\rangle\langle\psi^\prime_{k^\prime}|$ be another pure state decomposition of $\tau$,
and $\sum_h r_h |\phi_h\rangle\langle\phi_h|$ be a fixed pure state decomposition of $\rho_o-\tau\geq0$. Then,
$\sum_k q_k |\psi_k\rangle\langle\psi_k|+ \sum_h r_h |\phi_h\rangle\langle\phi_h|$ and
$\sum_{k^\prime} q_{k^\prime} |\psi^\prime_{k^\prime}\rangle\langle\psi^\prime_{k^\prime}|+\sum_h r_h |\phi_h\rangle\langle\phi_h|$
are two pure state decompositions of $\rho_o$, hence the equality $F_c(\rho_o)=F_a(\rho_o)$ implies
$\sum_k q_k F(|\psi_k\rangle)=\sum_{k^\prime} q_{k^\prime} F(|\psi^\prime_{k^\prime}\rangle)$
by definition of $F_a$ and $F_c$. Clearly, the maps $\tilde{F}$ and $F$ coincide on pure states, and moreover $\tilde{F}(0)=0$. Further, we claim that
\begin{eqnarray}
\tilde{F}(t_1\tau_1+t_2\tau_2)=t_1 \tilde{F}(\tau_1) + t_2 \tilde{F}(\tau_2)
\end{eqnarray}
{for all} $0\leq \tau_i \leq \rho_o$ and $ t_i\geq0$ with $t_1+t_2=1$.
Indeed, this follows by taking any pure state decompositions of $\tau_1$ and $\tau_2$ into weighted sums of pure states and applying (\ref{r1}) to both sides of the equation.

We can now define a functional on the space $S(\rho_o)$ of all self-adjoint operators with range in $R(\rho_o)$ as follows:  $p(H)=\inf (k_+\tilde{F}(\tau_+)-k_-\tilde{F}(\tau_-))$ where the infimum is taken over all nonnegative real numbers $k_+$ and $k_-$ and all density matrices $\tau_+$ and $\tau_-$ whose range is contained in $R(\rho_o)$ and for which $k_+\tau_+-k_-\tau_-=H$.  It is easy to verify that $p(H_1+H_2)\le p(H_1)+p(H_2)$ and $p(kH)=kp(H)$ for all nonnegative $k$.  Hence $p$ is a sublinear functional  on the space of all self-adjoint operators with range in $R(\rho_o)$. Note also that if $\rho$ is any density matrix with range in $R(\rho_o)$, we can see that $p(\rho)\le \tilde{F}(\rho)$ by choosing $k_+=1$, $k_-=0$, $\tau_{+}=\rho$ and $\tau_{-}$ to be any density matrix.  By choosing $k_+=0$, $k_-=1$, $\tau_{+}$  to be any density matrix and $\tau_{-}=\rho$, we can see that $p(-\rho)\le -\tilde{F}(\rho)$.

By the classical Hahn-Banach theorem, there exists a linear functional $L(H)$ on $S(\rho_o)$ such that $L(H)\le p(H)$ for all $H\in S(\rho_o)$.  Now if $\rho$ is any density matrix with range in $R(\rho_o)$, we get $L(\rho)\le p(\rho)\le \tilde{F}(\rho)$. We also get $-L(\rho)=L(-\rho)\le p(-\rho)\le -\tilde{F}(\rho)$ which after driving by minus one gives us $L(\rho)\ge \tilde{F}(\rho)$.  Combining our inequalities we get $\tilde{F}(\rho)=L(\rho)$.  Thus, there exists a nonnegative linear operator $Q:\ R(\rho_o)\rightarrow R(\rho_o)$ such that $\tilde{F}(\rho)=tr(Q\rho)$ for all states $\rho$ with $R(\rho)\subseteq R(\rho_o)$, which concludes the proof.

{\sf Proof of Corollary \ref{cor coh}}
For full rank quantum states $\rho$, if $C_c(\rho)=C_a(\rho)$, then there is a nonnegative linear operator $Q$ such that $C_f(|\psi\rangle)=\langle \psi |Q| \psi \rangle$ for all pure states in $R(\rho)=\mathcal{H}$. Since $C_f(|i\rangle)=\langle i|Q|i\rangle=0$ for all incoherent pure states $\{|i\rangle\langle i|\}_{i=0}^{n-1}$ in $\mathcal{H}$, $Q$ is a zero operator, which contradicts to $f \neq 0$.

\newpage
\bigskip
\noindent{\sf Acknowledgements}

\noindent The authors appreciate many useful suggestions and comments by the
anonymous referees. Ming-Jing Zhao thanks the Department of Mathematics and Statistics, University of Guelph, Canada for hospitality. Ming-Jing Zhao is supported by the China Scholarship Council (Grant No. 201808110022) and Qin Xin Talents Cultivation Program, Beijing Information Science and Technology University. Rajesh Pereira was supported by an NSERC Discovery grant (Grant No. 400550). Shao-Ming Fei is supported by the NSF of China (Grant No. 12075159), Beijing Municipal Commission of Education (KZ201810028042), Beijing Natural Science Foundation (Z190005), Shenzhen Institute for Quantum Science and Engineering, Southern University of Science and Technology (Grant No.SIQSE202005), and Academy for Multidisciplinary Studies, Capital Normal University.

\bigskip
\noindent{\sf Author contributions}

\noindent  All authors wrote and reviewed the manuscript.

\bigskip
\noindent{\sf Additional Information}

\noindent Competing Financial Interests: The authors declare no competing financial interests.

\end{document}